\documentclass[aps,prd,twocolumn,groupedaddress,amssymb,eqsecnum,showpacs,epsfig]{revtex4}
\usepackage{graphicx}
\usepackage{bm}
\usepackage{dcolumn}
\usepackage{amsmath}
\def \lleq {\lower0.9ex\hbox{ $\buildrel < \over \sim$} ~}
\def \ggeq {\lower0.9ex\hbox{ $\buildrel > \over \sim$} ~}

\def \beq  {\begin{equation}}
\def \eeq  {\end{equation}}
\def \ber  {\begin{eqnarray}}
\def \eer  {\end{eqnarray}}

\def\apj{{Astroph.\@ J.\ }}

\def\aj{{Astron.\@ J.\ }}

\begin{document}
\newcommand{\newc}{\newcommand}

\newc{\be}{\begin{equation}}
\newc{\ee}{\end{equation}}
\newc{\ba}{\begin{eqnarray}}
\newc{\ea}{\end{eqnarray}}
\newc{\bea}{\begin{eqnarray*}}
\newc{\eea}{\end{eqnarray*}}
\newc{\D}{\partial}
\newc{\ie}{{\it i.e.} }
\newc{\eg}{{\it e.g.} }
\newc{\etc}{{\it etc.} }
\newc{\etal}{{\it et al.}}
\newcommand{\nn}{\nonumber}
\newc{\ra}{\rightarrow}
\newc{\lra}{\leftrightarrow}
\newc{\lsim}{\buildrel{<}\over{\sim}}
\newc{\gsim}{\buildrel{>}\over{\sim}}
\title{Constraints on linear-negative potentials in quintessence and phantom models from recent supernova data}
\author{L. Perivolaropoulos}
\affiliation{Department of
Physics, University of Ioannina, Greece}
\date{\today}

\begin{abstract}
We study quintessence and phantom field theory models based on
linear-negative potentials of the form $V(\phi)=s\; \phi$. We
investigate the predicted redshift dependence of the equation of
state parameter $w(z$ for a wide range of slopes $s$ in both
quintessence and phantom models. We use the gold dataset of 157
SnIa and place constraints on the allowed range of slopes $s$. We
find $s=0\pm 1.6$ for quintessence and $s=\pm 0.7\pm 1$ for
phantom models (the range is at the $2\sigma$ level and the units
of $s$ are in $\sqrt{3}M_p H_0^2\simeq 10^{-38}eV^3$ where $M_p$
is the Planck mass). In both cases the best fit is very close to
$s\simeq 0$ corresponding to a cosmological constant. We also show
that specific model independent parametrizations of $w(z)$ which
allow crossing of the phantom divide line $w=-1$ (hereafter PDL)
provide significantly better fits to the data. Unfortunately such
crossings are not allowed in any phantom or quintessence single
field model minimally coupled to gravity. Mixed models (coupled
phantom-quintessence fields) can in principle lead to a $w(z)$
crossing the PDL but a preliminary investigation indicates that
this does not happen for natural initial conditions.
\end{abstract}
\maketitle

\section{Introduction}
Recent observations have indicated  that the universe has entered
a phase of accelerating expansion (the scale factor obeys ${\ddot
a}>0$) and that the total amount of clustered matter in the
universe is not sufficient for its small average spatial
curvature. This converging observational evidence comes from a
diverse set of cosmological data which includes observations of
type Ia supernovae \cite{snobs,Riess:2004nr}, large scale redshift
surveys \cite{lss} and measurements of the cosmic microwave
background (CMB) temperature fluctuations spectrum \cite{cmb}. The
observed accelerating expansion and flatness of the universe,
requires either a modified theory of gravity\cite{modgrav} or, in
the context of standard general relativity, the existence of a
smooth energy component with negative pressure termed `dark
energy'\cite{dark energy}. This component is usually described by
an equation of state parameter $w\equiv{p\over \rho}$ (the ratio
of the homogeneous dark energy pressure $p$ over the energy
density $\rho$). For cosmic acceleration, a value of $w<-{1\over
3}$ is required as indicated by the Friedmann equation \be {{\ddot
a}\over a}=-{{4\pi G}\over 3}(\rho +3p) \label{fried}\ee Even
though the cosmological constant remains a viable candidate for
dark energy, current observational bounds \cite{phant-obs2,nesper}
on the value of the dark energy equation of state parameter
$w(t_0)$ at the present time $t_0$ (corresponding to a redshift
$z=0$) yield $w(z=0)\leq -1$ with $\frac{dw}{dz}|_{z=0}>0$ at best
fit.

The role of dark energy can be played by any physical field with
positive energy and negative pressure which violates the strong
energy condition $\rho+3p>0$ ($w>-{1\over 3}$). Quintessence
scalar fields\cite{quintess} with positive kinetic term
($-1<w<-{1\over 3}$) violate the strong energy condition but not
the dominant energy condition $\rho + p>0$. Their energy density
scales down with the cosmic expansion and so does the cosmic
acceleration rate. Phantom fields\cite{phantom} with negative
kinetic term ($w<-1$) violate the strong energy condition, the
dominant energy condition and maybe physically unstable. However,
they are also consistent with current cosmological data and
according to recent studies\cite{phant-obs2,nesper} they maybe
favored over their quintessence counterparts.

Homogeneous quintessence or phantom scalar fields are described by
Lagrangians of the form \be  {\cal L}= \pm \frac{1}{2} {\dot
\phi}^2 - V(\phi) \label{lag1} \ee where the upper (lower) sign
corresponds to a quintessence (phantom) field in equation
(\ref{lag1}) and in what follows. The corresponding equation of
state parameter is \be w=\frac{p}{\rho} = \frac{\pm \frac{1}{2}
{\dot \phi}^2 - V(\phi)}{\pm \frac{1}{2} {\dot \phi}^2 + V(\phi)}
\label{eqst1} \ee For quintessence (phantom) models with $V(\phi)
> 0$ ($V(\phi) < 0$) the parameter $w$ remains in the range $-1 <
w < 1 $. For an arbitrary sign of $V(\phi)$ the above restriction
does not apply but it is still impossible for $w$ to cross the
phantom divide line (hereafter PDL) $w=-1$ in a continous manner.
The reason is that for $w=-1$ a zero kinetic term $\pm {\dot
\phi}^2 $ is required and the continous transition from $w<-1$ to
$w>-1$ (or vice versa) would require a change of sign of the
kinetic term. The sign of this term however is fixed in both
quintessence and phantom models. This difficulty in crossing the
PDL $w=-1$ could play an important role in identifying the correct
model for dark energy in view of the fact that data favor $w\simeq
-1$ and furthermore parametrizations of $w(z)$ where the PDL is
crossed appear to be favored over the cosmological constant $w=-1$
(see section III and Refs \cite{phant-obs2,nesper}).

In view of the above described problem it is interesting to
consider the available quintessence and phantom scalar field
models and compare the consistency with data of the predicted
forms of $w(z)$ among themselves and with arbitrary
parametrizations of $w(z)$ that cross the PDL. This is the main
goal of the present study.

We focus on a particular class of scalar field potentials of the
form \be V(\phi) = s \; \phi \label{pot1} \ee where we have
followed Ref. \cite{Garriga:2003nm} and set $\phi =0$ at $V=0$. As
discussed in section II (see also Ref. \cite{Garriga:2003nm}) the
field may be assumed to be frozen (${\dot\phi}=0$) at early times
due to the large cosmic friction $H(t)$. It has been argued
\cite{Garriga:1999bf} that such a potential is favored by
anthropic principle considerations because galaxy formation is
possible only in regions where $V(\phi)$ is in a narrow range
around $V=0$ and in such a range any potential is well
approximated by a linear function. In addition such a potential
can provide a potential solution to the cosmic coincidence
problem\cite{Avelino:2004vy}.

For quintessence models the scalar field behavior has been studied
extensively
\cite{Dimopoulos:2003iy,Garriga:2003nm,Kallosh:2003bq,Wang:2004nm}
 and shown
to lead to a future collapse of the scale factor (termed 'cosmic
doomsday') due to the eventual evolution of the scalar field
towards negative values of the potential where the gravity of the
field is attractive. Such a doomsday however does not occur in the
corresponding phantom models because the scalar field moves
towards higher values of the potential where the field gravity is
repulsive and leads to faster acceleration of the scale factor
(see Figure 2 below) and eventually to a Big Rip singularity
\cite{Caldwell:2003vq}. Thus $w$ evolves towards values less than
$-1$ for phantom models. One of the goals of this paper is to
compare the consistency with SnIa data of this phantom behavior of
$w(z)$ with the corresponding consistency of the quintessence
behavior where $w$ evolves towards values larger than $-1$.

The structure of the paper is the following: In the next section
we solve numerically the field equations for phantom and
quintessence models coupled to the Friedman equation and derive
the cosmological evolution of the scalar field, the scale factor
and the equation of state parameter $w$ for several values of the
potential slope $s$. In section III we fit the derived Hubble
parameter to the SnIa Gold dataset \cite{Riess:2004nr} and obtain
constraints for the potential slope $s$ for both phantom and
quintessence models. The quality of fit of these models is also
compared to the quality of fit of arbitrary $w(z)$
parametrizations that can cross the PDL. Finally in section IV we
summarize our results and state the main questions that emerge
from them. In a preliminary effort to address some of these issues
we show the evolution of $w(z)$ in a mixed quintessence + phantom
model where the dark energy consists of a mixture of interacting
phantom and quintessence fields.

\section{Phantom and Quintessence Field Dynamics}
In order to study in some detail the scalar field dynamics, we
consider the coupled Friedman-Robertson-Walker (FRW) and the
scalar field equation \ba \frac{\ddot a}{a}&=&\mp \frac{1}{3M_p^2}
({\dot \phi}^2 + s \; \phi)-\frac{\Omega_{0m} H_0^2}{2 a^3} \label{fried1} \\
{\ddot \phi} &+& 3 \frac{\dot a}{a}{\dot \phi} - s=0 \label{scal1}
\ea where $M_p = (8\pi G)^{-1/2}$ is the Planck mass and we have
assumed a potential of the form \be V(\phi)=\mp s \; \phi
\label{pot2} \ee where the upper (lower) sign corresponds to
quintessence (phantom) models. By setting \ba H_0 t & \rightarrow
& t \nn \\ \frac{\phi}{\sqrt{3} M_p} & \rightarrow & \phi  \\
\frac{s}{\sqrt{3} M_p H_0^2} & \rightarrow & s \nn \ea equation
(\ref{fried1}) may be written in rescaled form as \be \frac{\ddot
a}{a}= \mp  ({\dot \phi}^2 + s \; \phi)-\frac{\Omega_{0m}}{2 a^3}
\label{fried2} \ee while the scalar field equation (\ref{scal1})
remains unchanged. It is now straightforward to solve the system
numerically \cite{mathfile} using the following initial conditions
$(t\rightarrow t_i \simeq 0)$ \ba
a(t_i)&=&(\frac{9\Omega_{0m}}{4})^{1/3} \; t_i^{2/3} \nn \\
{\dot \phi}(t_i)&=&0 \\
\phi(t_i)&=&\phi_i \nn \ea since the universe is matter dominated
at early times and an inflationary phase would redshift the
gradient and velocity of the scalar field while the large cosmic
friction $\frac{\dot a}{a}$ would freeze it at early times after
inflation. The value of $\phi_i$ is chosen for each value of the
slope $s$ such that $\Omega_{0\phi}=\pm {\dot \phi}^2(t_0) +
V(\phi(t_0))=1-\Omega_{0m}$ at the present time $t_0$ (defined by
$a(t_0)=H(t_0)=1$). In what follows we have assumed a prior of
$\Omega_{0m}=0.3$. According to the numerical solution the scalar
field is almost frozen at early times (when matter dominates) due
to the large cosmic friction $H(t) \simeq \frac{2}{3t} $. At
approximately the present time when the matter density drops and
the field potential begins to dominate, the lower friction allows
the field to move down (up) the potential for quintessence
(phantom) models (see Figure 1).

\begin{figure}[h]
\centering
\includegraphics[bb=70 70 400 580,width=6.7cm,height=8cm,angle=-90]{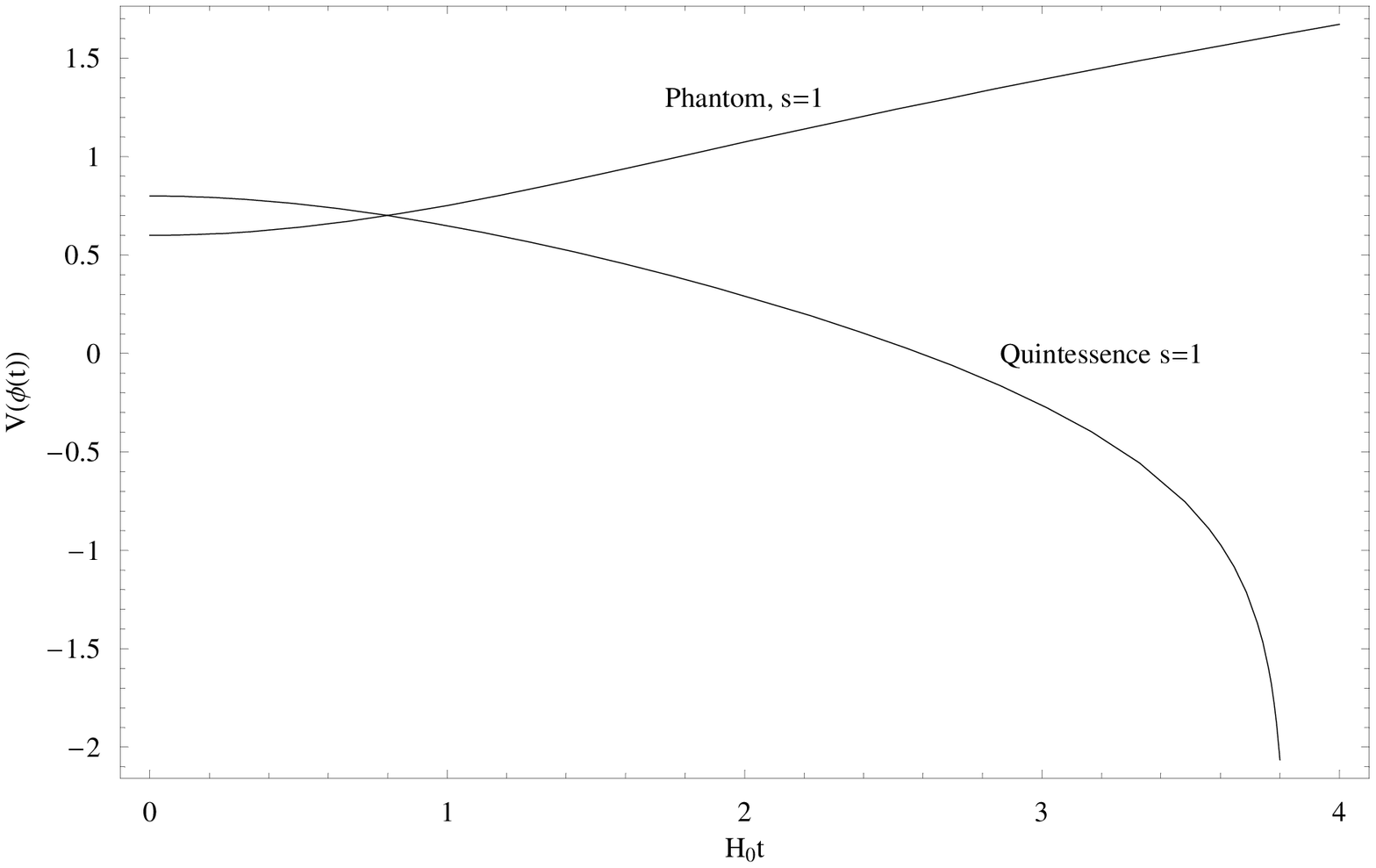}
\caption{The potential energy evolution for quintessence and
phantom models with linear potential of slope $s=1$. In this plot
$t_0=0.96$.} \label{fig1}
\end{figure}
When the potential energy dominates, the universe enters the
present accelerating phase. As the field moves down (up), the
potential energy becomes negative (more positive), the field
gravity becomes attractive (more repulsive) and the scale factor
begins to decelerate again (accelerate more rapidly) until the
universe ends with a Big Crunch (Big Rip) (see Figure 2).
\begin{figure}[h]
\centering
\includegraphics[bb=70 100 450 650,width=6.7cm,height=8cm,angle=-90]{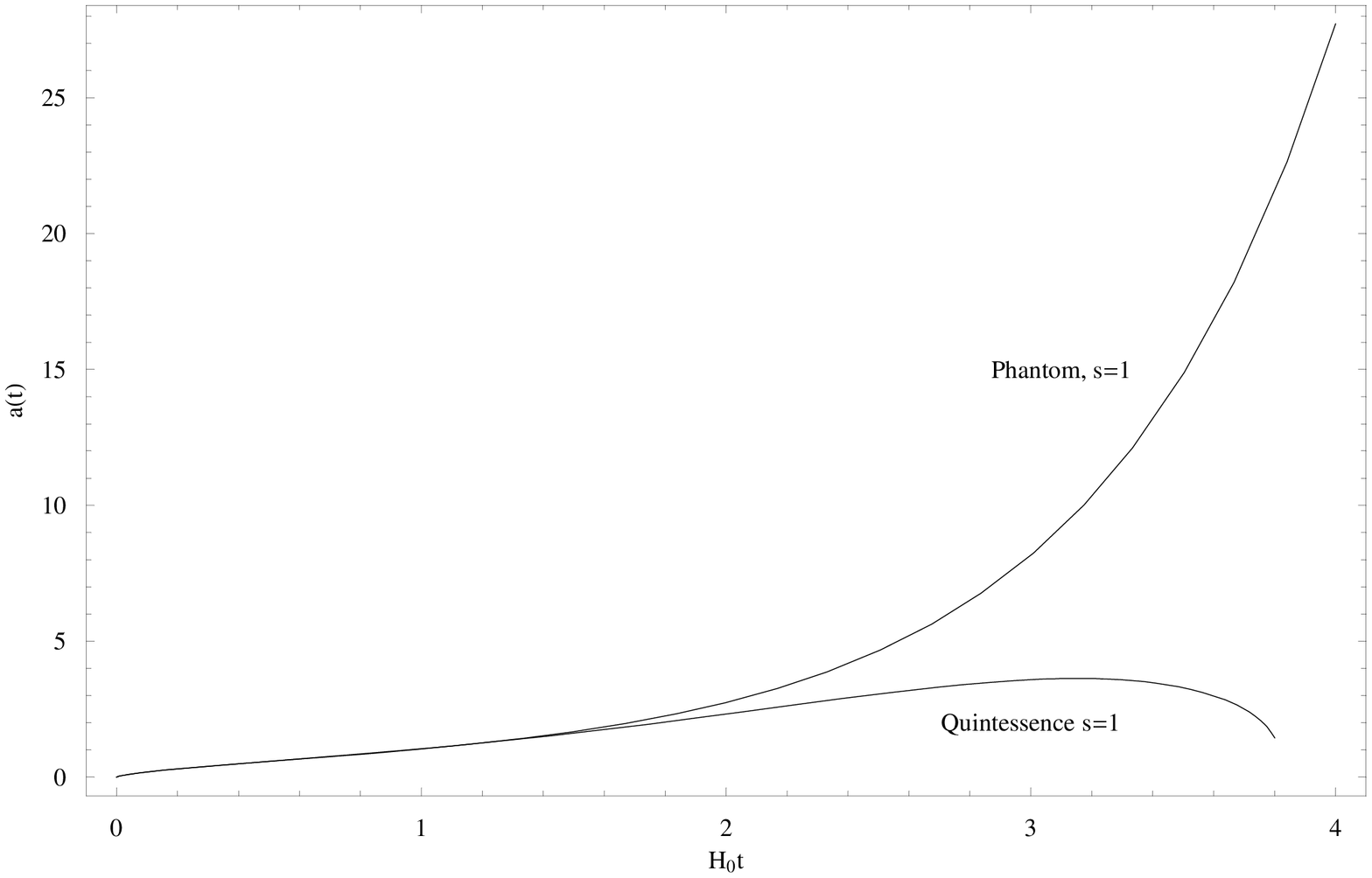}
\caption{The scale factor evolution for representative ($s=1$)
quintessence and phantom models with linear potential. The present
time corresponds to $t_0=0.96$ as in Figure 1.} \label{fig2}
\end{figure}
Using the numerical solution of the system
(\ref{fried2})-(\ref{scal1}) we can also evaluate the redshift
dependence of the equation of state parameter \be
w(z)=\frac{p}{\rho}=\frac{\pm \frac{1}{2} {\dot \phi}^2 -
V(\phi)}{\pm \frac{1}{2} {\dot \phi}^2 + V(\phi)}=\frac{
\frac{1}{2} {\dot \phi}^2 + s\; \phi}{ \frac{1}{2} {\dot \phi}^2 -
s\; \phi} \ee which is shown in Figure 3 in the redshift range
$0\leq z \leq 2$ for both quintessence and phantom models and for
several values of the slope $s$.

\begin{figure}[h]
\centering
\includegraphics[bb=70 100 510 720,width=6.7cm,height=8cm,angle=-90]{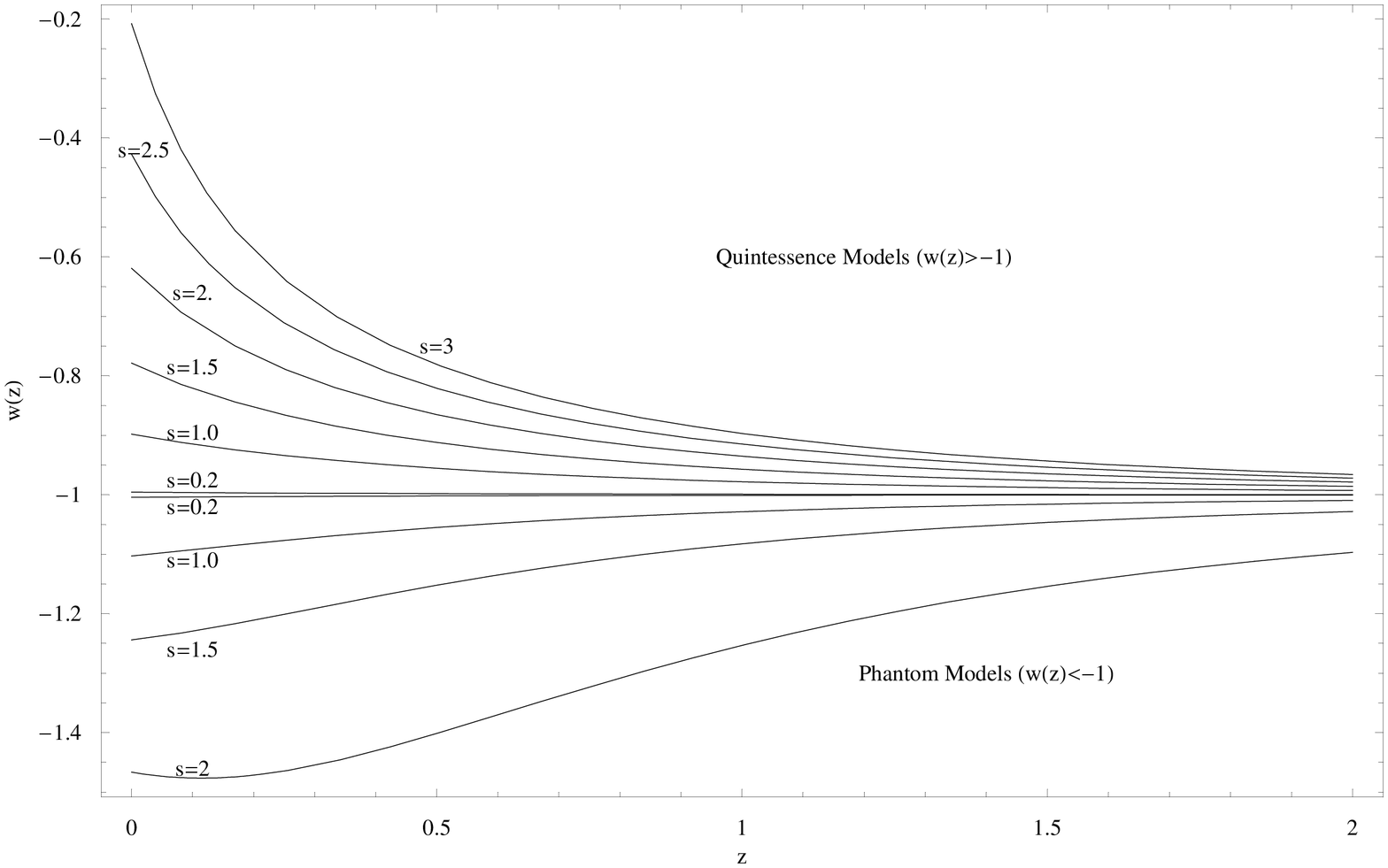}
\caption{The redshift evolution of the equation of state parameter
$w(z)$ for phantom and quintessence models and for several values
of the slope $s$.} \label{fig3}
\end{figure}

In the case of quintessence $w(z)$ has been evaluated in Ref.
\cite{Garriga:2003nm} (see also
\cite{Dimopoulos:2003iy,Kallosh:2003bq,Wang:2004nm}) with results
in good agreement with the corresponding results presented here.
As discussed in the introduction the PDL is not crossed for any
value of $s$. Instead, $w(z)$ evolves towards larger (smaller)
values of $-1$ for a quintessence (phantom) scalar field.

\section{Fit to the Gold Dataset}
Having solved numerically the rescaled system
(\ref{fried2}),(\ref{scal1}) for both quintessence and phantom
models, it is straightforward to obtain the corresponding Hubble
parameter $H(z;s)=\frac{\dot a}{a}(z;s)$ as a function of
redshift. This may now be used to obtain the corresponding Hubble
free luminosity distance \be D_L^{th} (z;s)= (1+z) \int_0^z
dz'\frac{H_0}{H(z';s)} \label{dlth1} \ee Using the maximum
likelihood technique \cite{press92} we can find the goodness of
fit to the corresponding observed $D_L^{obs} (z_i)$
$(i=1,...,157)$ coming from the SnIa of the Gold dataset
\cite{Riess:2004nr}. The observational data of the gold dataset
are presented as the apparent magnitudes $m(z)$ of the SnIa with
the corresponding redshifts $z$ and $1\sigma$ errors
$\sigma_{m(z)}$. The apparent magnitude is connected to $D_L(z)$
as \be m(z;s)={\bar M} (M,H_0) + 5 log_{10} (D_L (z;s))
\label{mdl} \ee where ${\bar M}$ is the magnitude zero point
offset and depends on the absolute magnitude $M$ and on the
present Hubble parameter $H_0$ as \be {\bar M} = M + 5
log_{10}(\frac{c\; H_0^{-1}}{Mpc}) + 25 \label{barm} \ee The
goodness of fit corresponding to any slope $s$ is determined by
the probability distribution of $s$ \ie \be P({\bar M}, s)= {\cal
N} e^{- \chi^2({\bar M},s)/2} \label{prob1} \ee where \be \chi^2
({\bar M},s)= \sum_{i=1}^{157} \frac{(m^{obs}(z_i) -
m^{th}(z_i;{\bar M},s))^2}{\sigma_{m^{obs}(z_i)}^2} \label{chi2}
\ee and ${\cal N}$ is a normalization factor. The parameter ${\bar
M}$ is a nuisance parameter and can be marginalized (integrated
out) leading to a new ${\bar \chi}^2$ defined as \be {\bar \chi}^2
=- 2 ln\int_{-\infty}^{+\infty} e^{-\chi^2/2} d{\bar M}
\label{barchi} \ee Using equations (\ref{chi2}) and (\ref{barchi})
it is straightforward to show (see Refs
\cite{DiPietro:2002cz,nesper}) that \be {\bar \chi}^2 (s) = \chi^2
({\bar M}=0,s)- \frac{B(s)^2}{C} + \ln(C/2\pi) \label{barchi2} \ee
where \ba B(s)&=&\sum_{i=1}^{157} \frac{(m^{obs}(z_i) - m^{th}(z_i
;{\bar
M}=0,s))}{\sigma_{m^{obs}(z_i)}^2} \label{bb} \\
C&=&\sum_{i=1}^{157}\frac{1}{\sigma_{m^{obs}(z_i)}^2 } \label{cc}
\ea Equivalent to marginalization is the minimization with respect
to ${\bar M}$. It is straightforward to show \cite{sfpriv} that
$\chi^2$ can be expanded in ${\bar M}$ as \be  \chi^2 (s) = \chi^2
({\bar M}=0,s)- 2 {\bar M} B  + {\bar M}^2 C  \label{chi2bm} \ee
which has a minimum for ${\bar M}=\frac{B}{C}$ at \be
\chi^2(s)=\chi^2 ({\bar M}=0,s)- \frac{B(s)^2}{C}
\label{chi2min1}\ee

Using (\ref{chi2min1}) we can find the best fit value of $s$
($s=s_0$) as the value that minimizes $\chi^2 (s)$ $(\chi^2
(s_0)=\chi_{min}^2)$. The $1\sigma$ error on $s$ is determined by
the relation \cite{press92} \be \Delta \chi_{1\sigma}^2 =
\chi^2(s_{1\sigma})-\chi_{min}^2 = 1 \label{dchi1s} \ee \ie $s$ is
in the range $[s_0, s_{1\sigma}]$ with $68\%$ probability.
Similarly the $2\sigma$ error ($95.4\%$ range) is determined by $
\Delta \chi_{2\sigma}^2  = 4 $ and the $3\sigma$ error ($99\%$
range) by $\Delta \chi_{3\sigma}^2  = 6.63$.

\begin{figure}[h]
\centering
\includegraphics[bb=10 100 410 650,width=6.7cm,height=8cm,angle=-90]{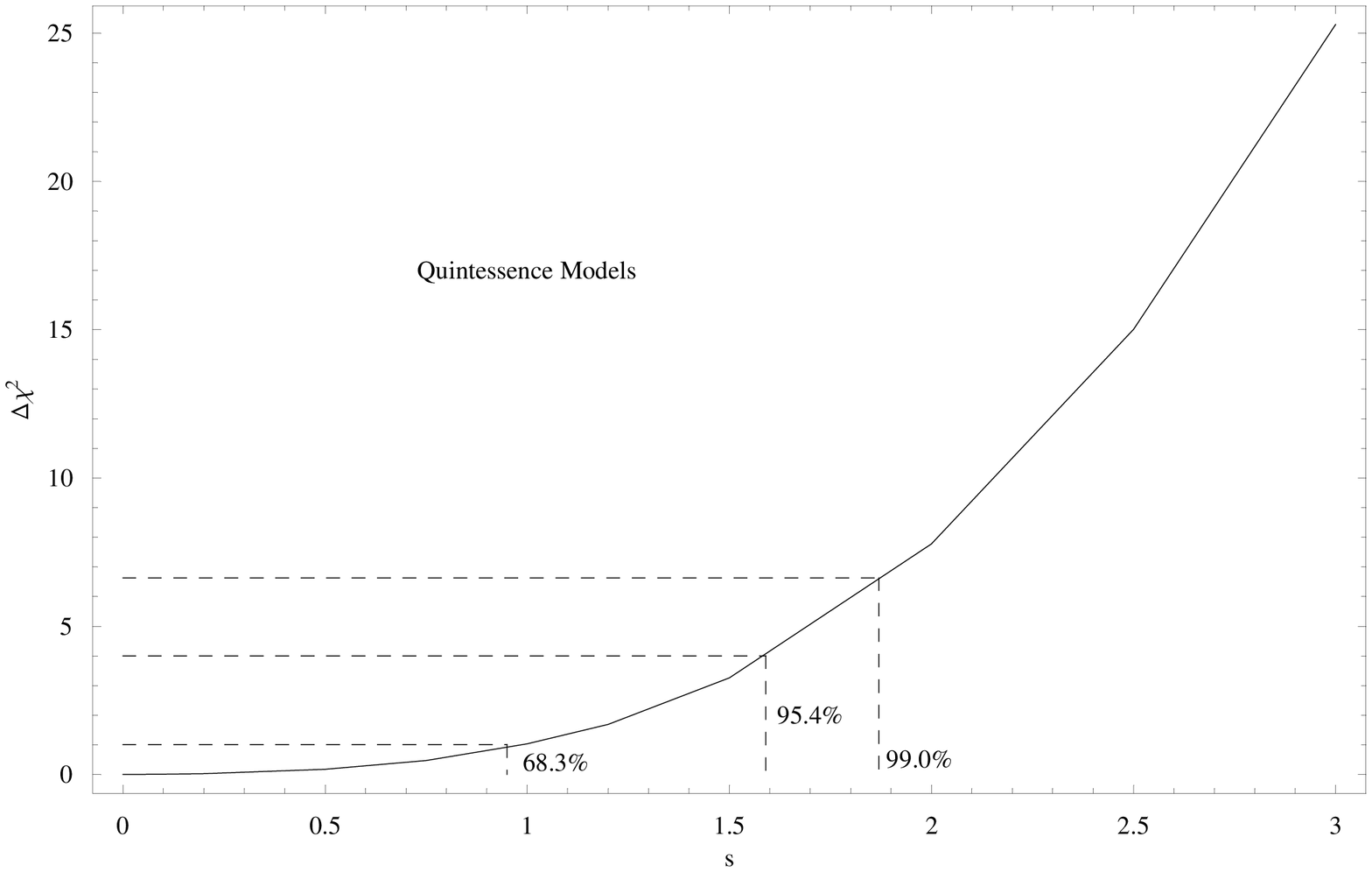}
\caption{The differences $\Delta \chi^2 (s)\equiv \chi^2 (s) -
\chi^2 (s\simeq 0)$ for quintessence models. The curve has been
sampled at $s=0,\; 0.1,\; 0.2,\; 0.5,\; 0.75,\; 1.0,\; 1.2,\;
1.5,\; 2.0,\; 2.5,\; 3.0$ and the corresponding points have been
joined.} \label{fig4}
\end{figure}

Figures 4 and 5 show plots of the differences $\Delta \chi^2
(s)\equiv \chi^2 (s) - \chi^2 (s\simeq 0)$ with respect to the
cosmological constant $(\chi^2 (s\simeq 0)=177.1)$ for
quintessence and phantom models with the $1\sigma$, $2\sigma$ and
$3\sigma$ ranges marked by dashed lines.

\begin{figure}[h]
\centering
\includegraphics[bb=40 100 450 650,width=6.7cm,height=8cm,angle=-90]{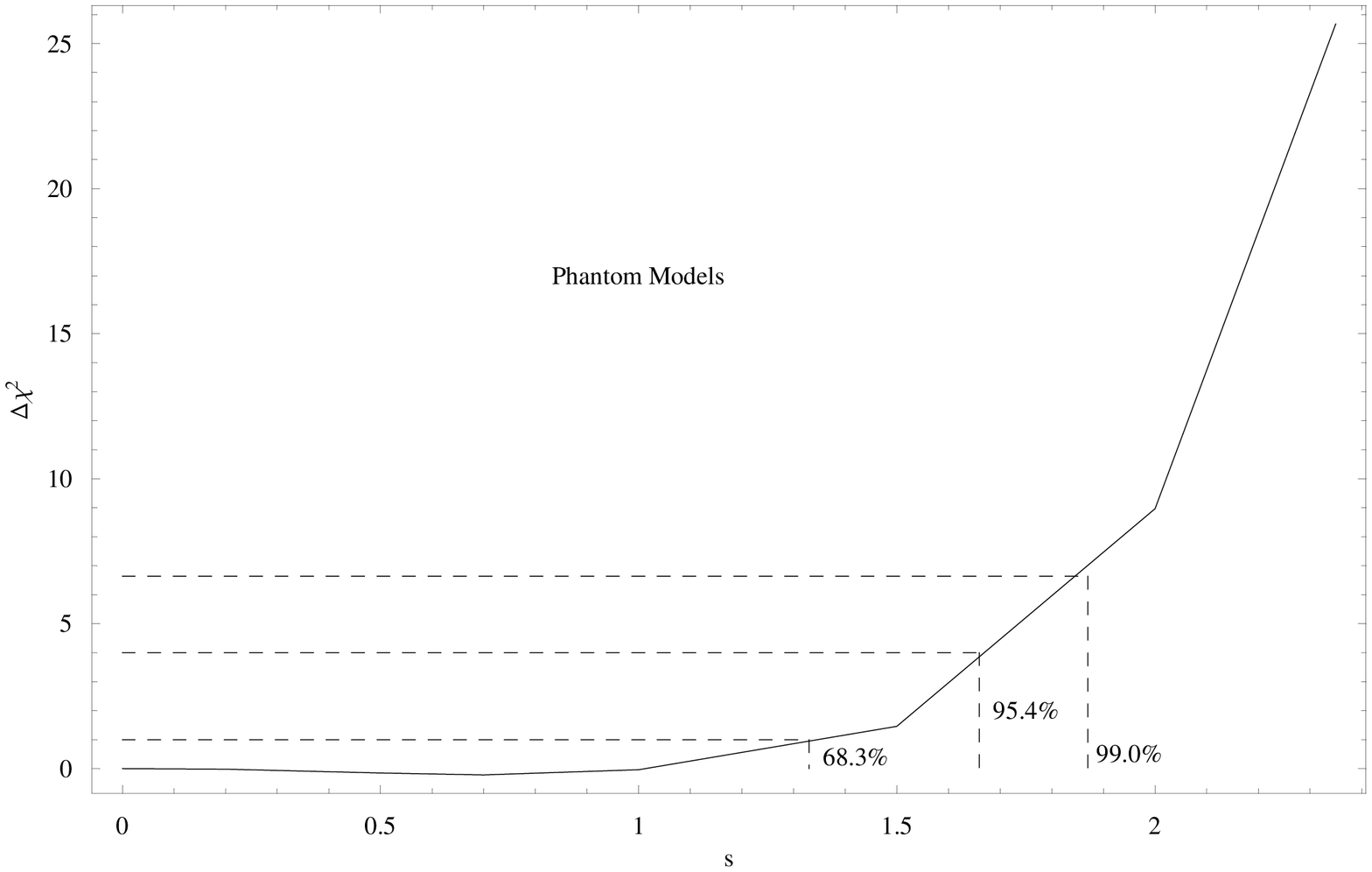}
\caption{The differences $\Delta \chi^2=\chi^2 (s) - \chi^2
(s\simeq 0)$ for phantom models. The curve has been sampled at
$s=0,\; 0.2,\; 0.5,\; 0.7,\; 1.0,\; 1.5,\; 2.0,\; 2.35$ and the
corresponding points have been joined. } \label{fig5}
\end{figure}

From Figures 4 and 5 it is clear that for both phantom and
quintessence models the best fit is obtained for $s\simeq 0$
corresponding to a cosmological constant. For phantom models
however the fit in the range $0<s<1$ is almost degenerate (all
values of $s$ in this range give essentially equally good fit).
The $2\sigma$ range for quintessence is $s\simeq 0 \pm 1.6$ while
for phantom fields the corresponding range is $s\simeq \pm 0.7 \pm
1$ (the best fit is at $s\simeq 0.7$ and the symmetry of the
evolution with respect to $+s \rightarrow -s$ has been imposed).
The value of $\chi^2$ at best fit is $\chi_{min}^2 = \chi^2
(s\simeq 0)=177.1$ for quintessence (identical to the value
corresponding to a cosmological constant with
$\Omega_{\Lambda}=0.7$) and $\chi_{min}^2=\chi^2(s\simeq
0.7)=176.9$ for phantom fields. Clearly both classes of models can
not provide better fits than the cosmological constant $\Lambda
CDM$.

We are thus faced with the following question: 'What are the
particular features required by $w(z)$ for better fits to the SnIa
data?'.  To address this question we can use arbitrary
parametrizations of $w(z)$ and identify the forms of $w(z)$ that
best fit the data. This task has been undertaken by several
authors \cite{phant-obs2,nesper} and the best fit forms of $w(z)$
found had the following common properties:
\begin{itemize}
\item The value of $w(z=0)$ at best fit was found to be in the
range $-1>w(z=0)>-2$ \item  The function $w(z)$ at best fit was
found to cross the PDL from below at least once with
$\frac{dw}{dz}>0$ in the range $0<z<1$.
\end{itemize}
This is demonstrated in Figure 6 where we plot $w(z)$ for two
representatives of the field theory models studied here
(quintessence with $s=2$ and phantom with $s=1.5$) superimposed
with $w(z)$ for the best fits of two arbitrary parametrizations.
The parametrizations considered are the following
\begin{itemize}
\item A linear ansatz \be w(z)=w_0+w_1 \; z \label{lineanz} \ee
Using the equation\cite{nesper} \be \label{wz3}
w(z)={{p_{DE}(z)}\over {\rho_{DE}(z)}}={{{2\over 3} (1+z) {{d \ln
H}\over {dz}}-1} \over {1-({{H_0}\over H})^2 \Omega_{0m} (1+z)^3}}
\ee we can obtain the Hubble parameter $H(z)$ corresponding to the
$w(z)$ of (\ref{lineanz}) as \ba \label{lin} && H^2 (z)=H_0^2
[\Omega_{0m} (1+z)^3 + \nn \\ &&(1-\Omega_{0m})
(1+z)^{3(1+w_0-w_1)} \; e^{3 w_1 z}] \ea This can now be used to
obtain $D_L^{th}(z;w_0,w_1)$ from equation (\ref{dlth1}) and
minimize the $\chi^2$ obtained from the Gold
dataset\cite{Riess:2004nr}. Using the Gold dataset, the best fit
parameter values for this ansatz are\cite{mathfile}
$(w_0,w_1)=(-1.4\pm 0.1,1.7\pm 0.4)$ giving $\chi^2 = 174.3$ at
the minimum (the errors are at the $1\sigma$ level). \item The
ansatz \be w(z)=w_0 + w_1 \frac{z}{1+z} \label{smanz}\ee which
varies between $w_0$ at $z=0$ and $w_0+w_1$ at $z\rightarrow
\infty$ with crossover at $z=1$ where the two values contribute
equally. The existence of such a crossover has the advantage that
observations near it apply to a reduced parameter phase space, and
hence the remaining parameter estimates are more
sensitive\cite{Linder:2002dt}. The Hubble parameter corresponding
to this ansatz is \ba \label{la} &H^2 (z)=H_0^2 [\Omega_{0m}
(1+z)^3 +& \nn \\ & (1-\Omega_{0m}) (1+z)^{3(1+w_0+w_1)} \; e^{3
w_1 ({1\over {1+z}} -1)}]& \ea The best fit parameter values for
this ansatz are\cite{mathfile} $(w_0,w_1)=(-1.6\pm 0.1,3.3\pm
0.5)$ giving $\chi^2 = 173.9$ at the minimum (the errors are at
the $1\sigma$ level).
\end{itemize}

These parametrizations were chosen for their relative simplicity
and for leading to fairly good fits to the data relatively to
other parametrizations (see also Ref. \cite{nesper}). As seen in
Figure 6 they share both of the properties refered above
($w(z=0)<-1$ and cross the PDL $w=-1$).

\begin{figure}[h]
\centering
\includegraphics[bb=40 80 450 630,width=6.7cm,height=8cm,angle=-90]{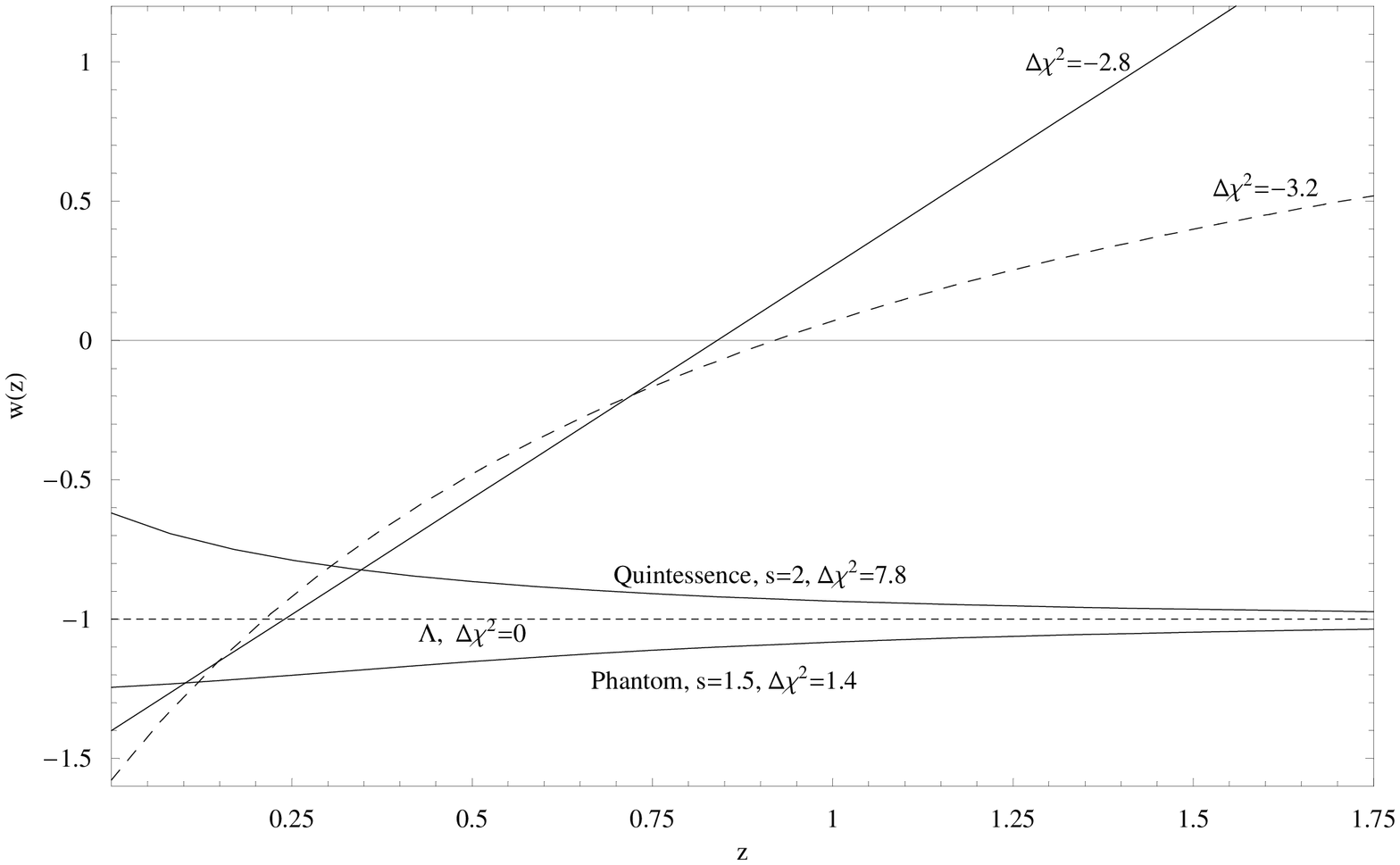}
\caption{The evolution of $w$ for two representatives of the field
theory models studied here superposed with two better fits
obtained by arbitrary parametrization: a linear ansatz
$w(z)=w_0+w_1 \; z$ and the ansatz $w(z)=w_0 + w_1 \frac{z}{1+z}$.
} \label{fig6}
\end{figure}
For each $w(z)$ we have evaluated $\Delta \chi^2$ with respect to
the cosmological constant $\chi_{\Lambda CDM}^2=177.1$. The field
theory models which do not cross the PDL have positive $\Delta
\chi^2$ and are therefore worse fits than $\Lambda CDM$. For
example a quintessence field with $s=2$ gives $\Delta \chi^2=7.8$
while a phantom field with $s=1.5$ gives $\Delta \chi^2=1.4$ (see
Figure 6). In contrast, the arbitrary parametrizations which cross
the PDL have negative $\Delta \chi^2$ and are therefore better
fits than $\Lambda CDM$. In particular for the linear ansatz
(\ref{lineanz}) we find $\Delta \chi^2=-2.8$ while for the
smoother ansatz of (\ref{smanz}) we find $\Delta \chi^2=-3.2$.
These differences mean that the point $(w_0,w_1)=(-1,0)$
corresponding to the cosmological constant from the viewpoint of
these parametrizations, is worse at more than $1\sigma$ from the
best fits obtained from these parametrizations. The problem with
such $w(z)$ parametrizations is that it seems to be highly
non-trivial to approximate their behavior using field theory
models (even exotic ones).

\section{Conclusion - Outlook}

We have shown that phantom and quintessence field theory models
have serious difficulty to exceed the quality of fit of a
cosmological constant to the SnIa Gold dataset. In contrast,
arbitrary parametrizations of the Hubble parameter and of $w(z)$
that cross the PDL can perform significantly better in fitting the
SnIa redshift data. We are thus faced with the question: 'Which
field theory models can mimic the behavior of the arbitrary
parametrizations crossing the PDL?'

As shown in the introduction this type of behavior can not be
achieved by either quintessence or phantom fields. In
principle\cite{Hu:2004kh} such behavior could be achieved by field
theories involving combinations of quintessence and phantom fields
(called 'quintom' in recent studies \cite{Guo:2004fq}). However,
preliminary investigations have demonstrated that mimicking the
$w(z)$ behavior indicated by the best fit parametrizations
requires significant fine tuning even for quintom field theories.
This can be demonstrated by studying the quintom Lagrangian \be
{\cal L}=\frac{1}{2}{\dot \phi}_1^2 - \frac{1}{2}{\dot \phi}_2^2 -
s\; \phi_1 - s\; \phi_2 - q \; \phi_1 \; \phi_2 \label{lquint} \ee
which leads to the rescaled dynamical equations \ba \frac{\ddot
a}{a}&=& -{\dot \phi}_1^2 +{\dot \phi}_2^2+ s \; (\phi_1 + \phi_2
) + q \; \phi_1 \; \phi_2
 -\frac{\Omega_{0m}}{2 a^3} \label{fried3}\\
 {\ddot \phi}_1 &+& 3 \frac{\dot a}{a}{\dot \phi}_1 + s + q \phi_2=0
\label{scal2}\\
{\ddot \phi}_2 &+& 3 \frac{\dot a}{a}{\dot \phi}_2 - s - q
\phi_1=0
\label{scal3}\ea

\begin{figure}[h]
\centering
\includegraphics[bb=90 80 500 700,width=6.7cm,height=8cm,angle=-90]{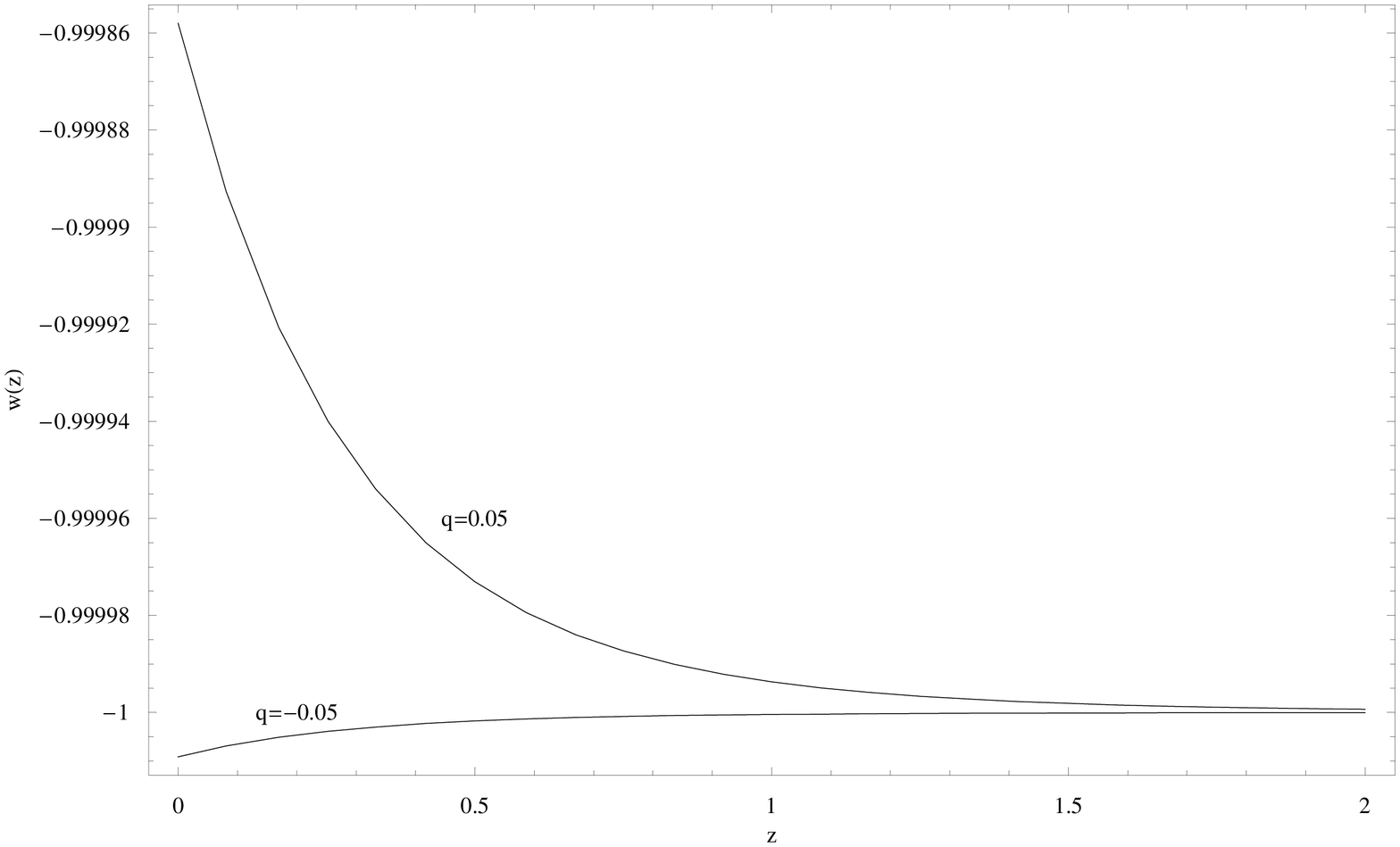}
\caption{The quintom model of the Lagrangian (\ref{lquint}) with
natural initial conditions and $s=0.2$ can mimic quintessence and
phantom models by changing the sign of the coupling $q$.}
\label{fig7}
\end{figure}

Even though we have not explored thoroughly the parameter space of
this quintom model we have shown that the natural initial
conditions used for the analysis of quintessence and phantom
systems presented above above do not lead to a crossing of the PDL
in the way indicated by the best fit parametrizations. As shown in
Figure 7 (obtained with $s=0.2$) we can simply mimic either the
quintessence or the phantom behavior of $w(z)$ by changing the
sign of the interaction coupling $q$.

Alternative field theory models that can in principle lead to a
$w(z)$ crossing the PDL with a single scalar field can be
constructed by considering generalizations of the kinetic term of
the Lagrangian in the form of
k-essence\cite{Armendariz-Picon:2000ah,Melchiorri:2002ux}. An
example of such models includes a Lagrangian with a field
dependent term multiplying the kinetic term in the form \be {\cal
L}=\frac{1}{2}f(\phi) {\dot \phi}^2 - V(\phi) \label{crpdl2} \ee
It has been shown however\cite{Vikman:2004dc} than even in such
models transitions through $w=-1$ are physically implausible
because they are either realized by a discrete set of trajectories
in phase space or they are unstable with respect to cosmological
perturbations.

\vspace{0.3cm} The Mathematica\cite{wolfram} file used for the
numerical analysis and the production of the figures of the paper
can be downloaded from \cite{mathfile} or sent by e-mail upon
request.

{\bf Acknowledgements:} I thank Stephane Fay for his help in
clarifying the relation between marginalization and minimization
with respect to ${\bar M}$ and John Rizos for useful discussions.
This work was supported by the General Secretariat for R\&D
research grant 'Pythagoras' no. 1705 (project 23).

\end{document}